\documentclass{aa}  
\usepackage{graphicx}
\usepackage{txfonts}
\usepackage{natbib}
\bibpunct{(}{)}{;}{a}{}{,}

\begin{document}
   \title{LABOCA observations of giant molecular clouds in the south west region of the Small Magellanic Cloud 
   }
   \titlerunning{LABOCA observation of GMCs in the SW of the SMC}

   \author{C. Bot \inst{1,}\inst{2}
          \and
          M. Rubio\inst{3}
          \and
          F. Boulanger\inst{4}
          \and
          M. Albrecht\inst{5,6}
          \and
          A. Leroy\inst{7}
          \and
          A. D. Bolatto\inst{8}
          \and
          F. Bertoldi\inst{5} 
          \and
          K. Gordon\inst{9}
          \and
          C. Engelbracht\inst{10}
          \and
          M. Block\inst{10}
          \and
          K. Misselt\inst{10}
          }

   \offprints{C. Bot}

   \institute{Universit\'e de Strasbourg, Observatoire Astronomiques de Strasbourg\\
              \and
              CNRS, Observatoire Astronomique de Strasbourg, F-67000 Strasbourg\\
              UMR7550, F-67000 Strasbourg, France\\
              \email{caroline.bot@astro.unistra.fr}
              \and
             Departamento de Astronomia, Universidad de Chile, Casilla 36-D, Santiago, Chile\\
         \and
             Institut d'Astrophysique Spatiale, Universit\'e Paris-Sud, F-91405, Orsay, France\\
             \and
             Argelander-Institut f\"ur Astronomie, Universit\"at Bonn, Germany\\
            \and
             Universidad Cat\'olica del Norte, Chile\\
             \and
             Max Planck Institute for Astronomy, Heidelberg, Germany\\
             \and
             University of Maryland, MD, USA\\
             \and
             Space Telescope Science Institute, 3700 San Martin Drive, Blatimore, MD21218, USA
             \and
             Steward Observatory, University of Arizona, Tucson, AZ 85721\\
             }

   \date{Received ...; accepted ...}
 
  \abstract
   {The amount of molecular gas is  a key for understanding the future star formation in a galaxy. Because H$_2$ is difficult to observe directly in dense and cold clouds, tracers like the CO molecule are used. However, at low metallicities especially, CO only traces the shielded interiors of the clouds. In this context, millimeter dust emission can be used as a tracer to unveil the total dense gas masses. However, the comparison of masses deduced from the continuum SIMBA 1.2~mm emission and virial masses (thought to trace the whole potential of the clouds) in a sample of giant molecular clouds in the Small Magellanic Cloud (SMC) by Rubio et al. (2004) and Bot et al. (2007) showed a discrepancy between these two quantities that is in need of an explanation.}
   {This study aims at better assessing possible uncertainties on the dust emission observed in the sample of giant molecular clouds from the SMC. It focuses on the mass comparison in the densest parts of the giant molecular clouds where CO is detected in order to confirm the mass discrepancy observed.}
   {New observations of the south-west region of the SMC were obtained with the LABOCA camera on the APEX telescope. All the giant molecular clouds previously observed in CO are detected and their emission at 870$\mu$m is compared to ancillary data. The different contributions to the sub-millimeter emission are estimated, as well as dust properties (temperatures, emissivities), in order to deduce molecular cloud masses precisely.}
   { The (sub-)millimeter emission observed in the giant molecular clouds in the south west region of the SMC is dominated by dust emission and masses are deduced for the part of each cloud where CO is detected and compared to the virial masses. The mass discrepancy between both methods is confirmed at 870$\mu$m with the LABOCA observations: the virial masses are on average 4 times smaller than the masses of dense gas deduced from dust emission, contrary to what is observed for equivalent clouds in our Galaxy. }
   { At present, the origin of this mass discrepancy in the SMC remains unkown. The direct interpretation of this effect is that the CO linewidth used to compute virial masses do not measure the full velocity distribution of the gas. Geometrical effects and uncertainties on the dust properties are also discussed.}


   \maketitle
%

\section{Introduction}

Star formation is observed to occur within molecular clouds. The star formation rate in a galaxy therefore depends on the amount of molecular gas available, but this quantity remains difficult to estimate precisely. In particular, at low metallicities, the relationship between H$_2$ and CO  (the most widely used tracer of molecular gas) remains unclear.

The Small Magellanic Cloud (SMC) is one of the closest and easiest to observe, low metallicity galaxies. Due to its proximity, molecular clouds in this galaxy can be resolved at various wavelength. Numerous molecular clouds in the Magellanic clouds were observed in CO lines \citep{IJL93,RLB93,RLB+93,RLB+96} and were found to have weak CO emission. If the CO molecule is tracing accurately the molecular gas, then the star formation per unit CO emission in the SMC is higher than in most galaxies. Such an effect is observed in other low metallicity dwarf galaxies \citep{Leroy:2006xy} and could be an "evolutionary effect" or the fact that CO observations largely underestimate the amount of molecular gas in these conditions. Indeed, at low metallicities, CO can be photodissociated and trace only the densest parts of the clouds, while H$_2$ molecules have the capacity to self-shield and would be present at larger radius than the CO molecule.

This "underestimation" by CO of the molecular content of SMC molecular clouds has been seen and studied mainly through the $X_{CO}=N_{H_2}/I_{CO}$ factor. Using CO data only, the velocity dispersion observed by the CO line width is thought to trace the gravitational potential of the cloud and therefore the total molecular content, which can be compared to the CO intensity in the same region. Analyzing NANTEN data for the whole SMC, \citet{Mizuno:2001uo} found $X_{CO}=(1-5)\times 10^{21} \mathrm{cm}^{-2} (\mathrm{K.km.s}^{-1})^{-1}$. \citet{Blitz:2007kx} corrected these data for sensitivity and resolution effects and found $X_{CO}=(1-1.5)\times 10^{21} \mathrm{cm}^{-2} (\mathrm{K.km.s}^{-1})^{-1}$. These values are systematically larger than the canonical galactic value $X_{CO}=2.3 \times 10^{20} \mathrm{cm}^{-2} (\mathrm{K.km.s}^{-1})^{-1}$ \citep{SBD+88}, supporting the idea that CO is present only in part of the molecular cloud. There is evidence for a dependence of the $X_{CO}$ factor with radius \citep{RLB93} and at the smallest resolved scales $X_{CO}$ determined for individual resolved clouds might become close to the galactic value \citep{BLI+03,Israel:2003yq,Bolatto:2008qr}.

To better understand the total molecular content and its relationship to CO emission, other tracer of the dense and molecular gas at low metallicities have been looked for. Dust continuum emission can be used to trace the dense and cold interstellar medium where the gas is molecular and it can therefore be used to unveil the total mass of dense gas in clouds. \citet{Isr97} was the first to introduce this method in the Magellanic Clouds, comparing IRAS and CO data, and he deduced higher H$_{2}$ masses than what was found using CO alone. More recently, \citet{Leroy:2007fk} attempted to spatially trace molecular hydrogen using far infrared emission observed with Spitzer data in the SMC. In dense regions, they observed large quantities of dust not associated to either H{\sc i} or CO emission. A study of the star forming region N83 confirms this result on smaller scale \citep{Leroy:2009rt}.
However, the dust emission in the far-infrared is sensitive to the dust temperature, leading to large uncertainties in the cloud masses. In the sub-millimeter and the millimeter range, dust emission is difficult to observe but depends only linearly on the dust temperature and is optically thin. \citet{RBR+04} observed for the first time with SIMBA on the SEST telescope, the 1.2~mm emission from dust in a quiescent molecular cloud in the SMC. The dense gas mass deduced were several times higher than the dynamical mass deduced from CO observations in this cloud. \citet{Bot:2007yq} extended this study with SIMBA to a larger sample of molecular clouds in the south-west region of the SMC. In all clouds, the gas mass deduced from the dust emission is systematically larger than the virial mass deduced from CO, even for conservative values of the dust emissivities and free-free contribution to the millimeter emission. All these studies come to agreement in the fact that molecular clouds in the SMC are more massive than what can be deduced from CO emission. This mass difference is understood in a scenario where CO traces only the densest parts of the molecular clouds.

However, in case of virial equilibrium, it has been assumed that the observed motions should balance the gravitational pressure and the dynamical masses obtained from molecular line observations should trace the total molecular masses \citep[e.g.][]{MRW88,Rosolowsky:2003uq,Du:2008kx,Dunham:2010fk}. The result from \citet{Bot:2007yq} where the virial masses in the SMC systematically underestimate the true mass of the clouds (as traced by the dust mass) remains unexpected. Partial support of the clouds by a magnetic field has been invoked to explain this mass discrepancy. However, these results need to be confirmed with observations in the sub-millimeter range in order to more certainly exclude any possible contamination to the fluxes (free-free emission, CO line contribution), as well as unknown instrumental effects and dust properties. This study focuses on the parts of the giant molecular clouds that are detected in CO. These regions represent only the highest density parts of the giant molecular clouds but are most appropriate for the comparison with CO and the check of a mass discrepancy.

We present new observations of the sample of molecular clouds in the south west region of the SMC taken with the LABOCA camera on the APEX telescope (section \ref{sec:data}). These observations at 870$\mu$m complement the SIMBA data at 1.2mm, CO and radio data, as well as 160$\mu$m Spitzer data from the combined S$^3$MC and SAGE-SMC surveys \citep{Bolatto:2007rc, Gordon:2009lq}. We build spectral energy distributions of each giant molecular cloud (GMC) detected both in the sub-millimeter and in CO and assess the possible contaminations and uncertainties(Sec. \ref{sec:origemi}). The dust emission in each cloud is then used to determine dense gas masses for each molecular cloud in the south-west region of the SMC. These mass estimates are compared to virial masses and confirm the mass discrepancy (Sec. \ref{sec:results}).

\section{Data and methods}\label{sec:data}

\subsection{The data}
The dust continuum observations were obtained using the Large APEX Bolometer Camera 
\citep[LABOCA,][]{Siringo:2009vn} on the Atacama Pathfinder Experiment \citep[APEX,][]{Gusten:2006kx} telescope in Chile. LABOCA is a bolometer array consisting of 295 receivers packed in a hexagonal structure resulting in a total field of view of 11\farcm 4. It has a central frequency of 345~GHz ($\sim 870 \mu$m) with a band width of 60~GHz and the beam-size (HPBW) was measured to be 19\farcs 2.

The $870\mu$m LABOCA observations were done during October 2007. The weather conditions were mostly good to excellent with a precipitable water vapor (PWV) content typically between 0.5 to 0.9~mm resulting in zenith opacities of the order of 0.2.   
The pointing was checked regularly on strong continuum sources and planets, and the pointing accuracy was found to be mostly better than 3\arcsec. The focus settings were determined from observations of planets typically once per night and during changes of the weather conditions, sunset and sunrise. To obtain the atmospheric opacity the sky emission was measured regularly about every 
two hours. Calibration factors were derived by observing Neptune and Uranus as well as
secondary calibrators. Relative offsets and relative gains (flatfield) of the individual bolometer channels 
are determined about once per month and are provided by the APEX staff.   
The scientific mapping was carried out in spiral raster mode with the raster pattern alternating between 2$\times$2 and 3$\times$3 spirals to achieve an optimal sampling of the region of interest.
Nine positions raster maps were done in a 140" spiral at 1"/s speed and 60$^o$ inclination. Raster lengths were varied in $x$ and $y$ directions between 1000" and 1200", and the spacings between 500" and 600". Dithering of 100" to 120" from the center position was performed. Additional observations of the central $6'\times 6'$ region in a four by four points raster were performed to increase the signal to noise ratio. In total we obtained  101 maps and a pure on-source integration time of 8.9~hours.    
 The final rms achieved was 8mJy/beam.

The LABOCA data were reduced using the reduction package BoA \footnote{http://www.astro.uni-bonn.de/boawiki/Boa} (Schuller et al. in prep). The processing of the raw time series, i.e. the time-ordered data stream (down-sampled to 25~Hz) of each channel and scan, consists of the following steps:  calibration correction by applying a linearly interpolated calibration factor; opacity correction by applying the linearly interpolated opacity at the  elevation of the scan; correction for temperature drifts due to the cryosystem using two bolometers  that have been sealed to block the sky signal for this purpose; flat fielding by applying the relative bolometer gains; conversion from counts to Jansky; flagging of bad channels.  In addition to the known dead or noisy channels, bad channels were identified as those having an rms of their time series a given ratio $r$ higher or $1/r$ lower than the median rms of all channels.  
The further reduction applied (in part on several occasions) the following tasks: correlated noise removal on the full array; correlated noise removal on groups sharing the same amplifier box and the same wiring; baseline subtraction; despiking; flagging of data outside suitable telescope scanning velocity and/or  acceleration limits;  flagging of bad channels. Each reduced scan (i.e. raster pattern) was then gridded into a weighted intensity map and a corresponding weight map. The weights of the data points contributing to a certain pixel of the intensity map were determined as $1/\sigma^2$ where $\sigma$ denotes the rms of the reduced time series of the corresponding channel and subscan. Individual maps were then coadded, again noise-weighted, to build the final intensity map and the corresponding weight map, which in turn allows to retrieve the rms for each pixel and to construct a signal-to-noise (S/N) map.   

The described data reduction is affected by the presence of astronomical signals in the time series, which leads to a flux loss due the subtraction of correlated noise. In case of strong sources this is manifestly revealed by the appearance of areas of negative fluxes adjacent to the sources. Moreover, the source emission wrongly contributes to the determination of the rms and  weights.  To minimise these effects we applied an iterative approach by subsequently improving a  model of the flux distribution of the astronomical source. 

In a first step a map was produced by a ``blind'' execution of the reduction as described above. A first source model was constructed from the resulting map by extracting all pixels above a given S/N threshold (typically 3) and setting the remaining pixels to 
zero. This model was converted to a time series for each channel and subtracted from the databefore any baseline subtraction, despiking and correlated noise suppression. After the weight determination and before the map was built, the model was re-introduced
into the time stream. The resulting map was used to extract a new source model. In subsequent iterations this process is repeated until the measured flux distribution converges. As a result, most of the faint extended emission should be recovered in the reduction. 

\begin{figure}
\includegraphics[width=0.4\textwidth,angle=90]{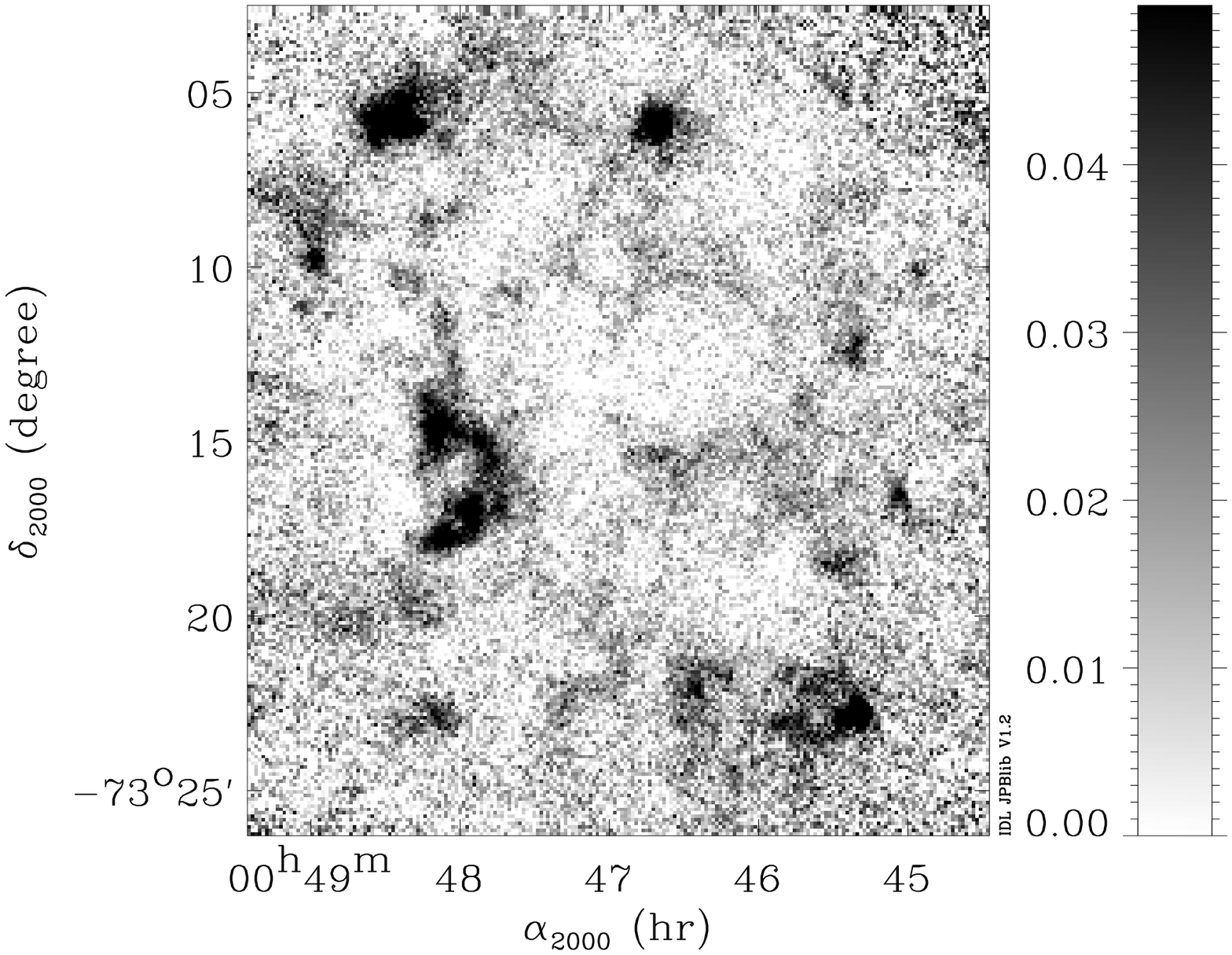}\\
\includegraphics[width=0.4\textwidth,angle=90]{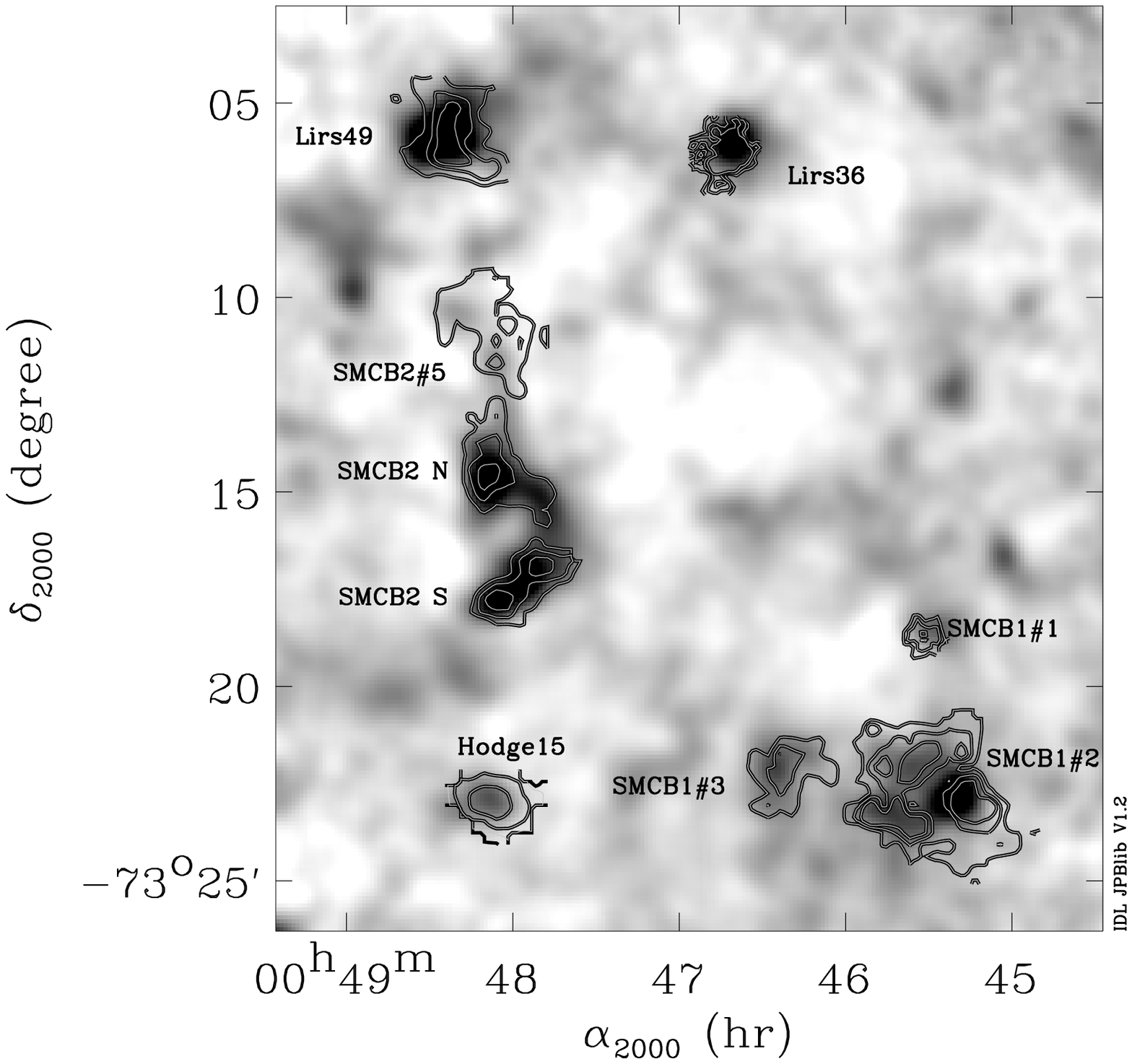}
 \includegraphics[width=0.4\textwidth,angle=90]{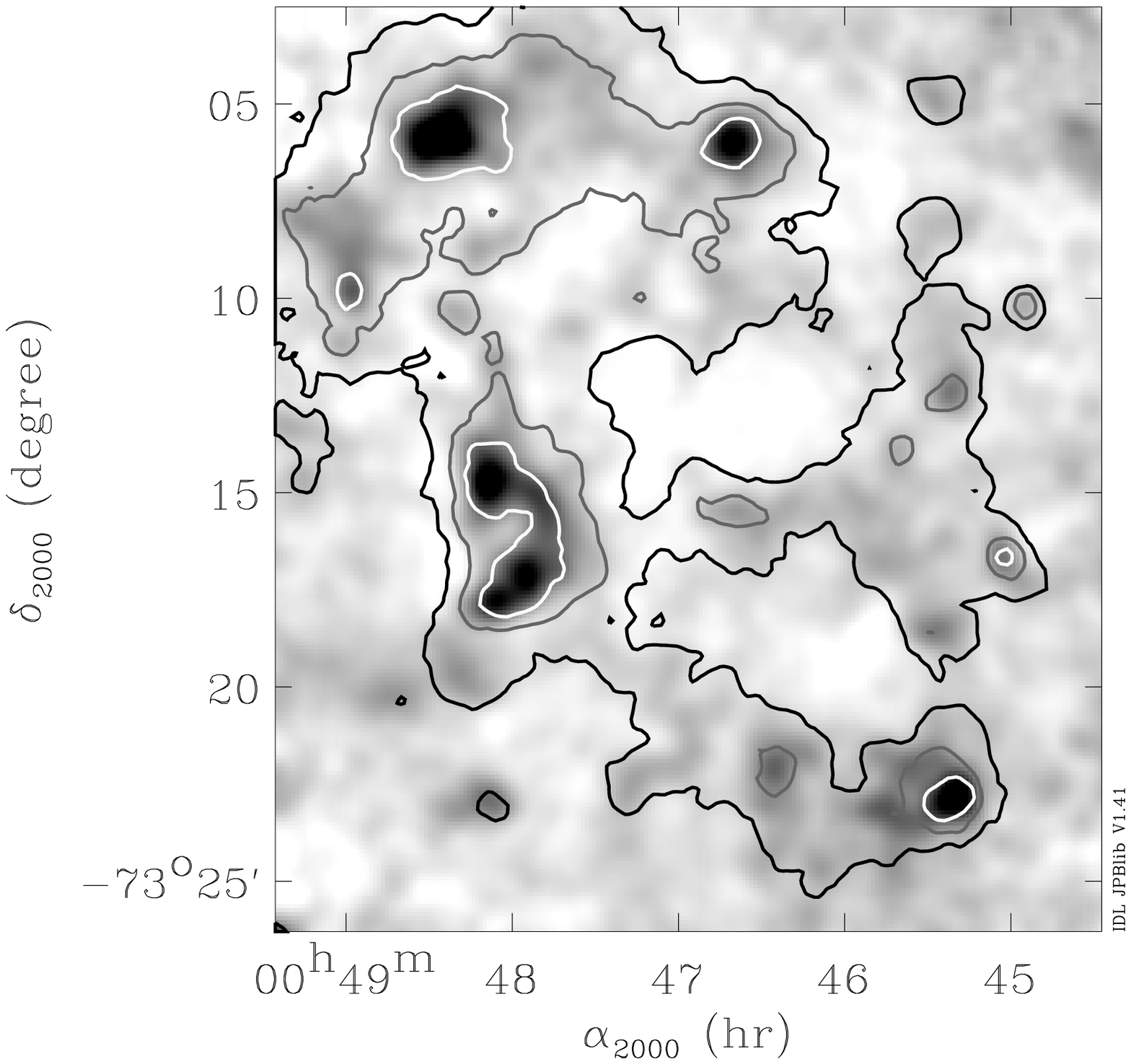}\\
\caption{Top image: LABOCA image in the south west region of the SMC (Units are in Jy/beam). Middle and bottom images: LABOCA images convolved to a $\sim 40$" resolution with CO and MIPS 160$\mu$m contours overlaid respectively. All the molecular clouds observed in CO are detected at 870$\mu$m. Extended emission is well recovered in the LABOCA map and the 870$\mu$m emission matches well the 160$\mu$m one. \label{fig1}}
\end{figure}

The 20'$\times$20' image of the south-west region of the SMC that was obtained, is presented in Fig. \ref{fig1}. It is the first sub-millimeter map of a large region in the Small Magellanic Cloud. This south-west region of the SMC has had most of its known molecular clouds mapped at a 43" resolution  and at hight sensitivity in the CO (J=1-0) line with SEST \citep{RLB93,RLB+93}. All the GMCs observed in CO lines with SEST (shown as contours in Fig. \ref{fig1}) are detected with the present LABOCA observation. There is also extended emission well outside observed CO peaks.  If the 870$\mu$m emission originates from dust, this result shows that there is much more dense gas than what is traced by current CO observations. Comparing with the MIPS 160$\mu$m map of the same region, we see that the spatial distribution of the 870$\mu$m emission is well correlated with the one observed in the far-infrared. A short analysis of the dust emission outside the observed CO is presented in Appendix \ref{appendb}.

The present paper compares the 870$\mu$m and CO data and is therefore restricted to the regions where CO has been observed and detected. The entities for which we deduce masses in the following and that are refered to as "clouds" are the densest parts of the molecular clouds, associated with CO emission. 

For comparison, we complement these data with the SIMBA observations at 1.2~mm previously studied \citep{Bot:2007yq} and with Spitzer MIPS observations at 160$\mu$m in the same region from the combined S$^3$MC and SAGE-SMC surveys \citep{Bolatto:2007rc, Gordon:2009lq}.  Furthermore, in order to estimate the contribution of free-free emission to the observed millimeter emission, we also use radio ATCA maps at 8.6, 4.8~GHz that were obtained from J. Dickel (http://www.phys.unm.edu/$\sim$johnd/).

\subsection{Flux estimates}

To estimate the LABOCA, SIMBA, MIPS and radio fluxes in a way that is comparable to CO, we do the following. All maps are convolved to the 43" CO(1-0) resolution and projected on the same sampling grid. Because we are interested in comparing the LABOCA, SIMBA, MIPS and CO fluxes, it is important to remove the extended emission in all maps in order to compare emission in the region detected with all tracers (the peaks of the GMCs). To do so, we begin by masking the region where CO is detected and fit in each map (LABOCA, radio, SIMBA, MIPS) a 2D plane to the emission outside the CO detected region. The fitted plane is then removed from the maps (in units of surface brightness) to leave only the peak of emission where CO was observed. The integrated fluxes are then computed in a region above a defined $W_{CO}$ threshold. This threshold is defined individually for each cloud so that the area of the cloud we sample is similar to the one that was used to determine the virial masses and CO luminosities of the clouds \citep{RLB+93}. This method enables us to analyze the emission coming from the dust associated with the CO emission only. It attempts to filter extended envelopes around the clouds -- uncertainties on this filtering process will be discussed in section \ref{sec:geom} -- 
 as well as possible emission associated with H{\sc i}.  Using an H{\sc i} map of the Small Magellanic Cloud \citep{SSD+99}, we estimated that the remaining H{\sc i} column densities associated to the peaks of the molecular clouds we study, after filtering, are less than a few percent of the total hydrogen column densities as determined from the dust millimeter emission. The fluxes obtained are summarized in Tab. \ref{tab1} and are used to build the spectral energy distribution for each cloud (c.f Fig. \ref{fig2}). 

\begin{table*}
\centering
\caption{Observed fluxes from the far-infrared to the radio for our sample of giant molecular clouds. The radii and velocity dispersions $\Delta V$ from \citet{RLB+93} (and \citet{LLD+94} for Hodge 15) used to compute the virial masses are also mentioned. Notes: (1) Association of the molecular clouds $[$RLB93$]$ SMC-B2 1,  $[$RLB93$]$ SMC-B2 2 and  $[$RLB93$]$ SMC-B2 6; (2)  Association of the molecular clouds $[$RLB93$]$ SMC-B2 3 and  $[$RLB93$]$ SMC-B2 4\label{tab1}. }
\begin{tabular}{l l l l l l l l l l}
name & SIMBAD name & Radius & $\Delta V$ &  $S_{160\mu m}$ &$S_{870 \mu m}$ &  $S_{1.2mm}$ & $S_{3cm}$ & $S_{6cm}$ \\
 & & pc & km.s$^{-1}$& Jy &mJy & mJy  &mJy & mJy  \\
\hline
LIRS49 & LI-SMC 49 & $16.7$ & $6.2$ & $16.1\pm 1.5$ & $1358\pm 113$ & $403\pm 66$ & $28\pm 3$ & $6\pm 3$  \\
LIRS36 & LI-SMC 36 & $18.6$ & $3.8$ & $7.2\pm 0.8$ & $1015\pm 104$ & $258\pm 45$ & $16\pm 4$ & $20\pm 3$  \\
SMCB1\#1 & $[$RLB93$]$ SMC-B1 1 & $13.8$ & $3.1$ &$1.6\pm 0.4$ & $234\pm 4$ & $83\pm 25$ & $<6$ &$<5$  \\
SMCB1\#2 & $[$RLB93$]$ SMC-B1 & $16.2$ & $4.9$ & $8.9\pm 2.0$ & $789\pm 146$ & $175\pm 76$ & $15\pm 3$ & $14\pm 2$  \\
SMCB1\#3 & $[$RLB93$]$ SMC-B1 3& $12.9$ & $3.2$ &$2.9\pm 1.4$ & $265\pm 84$ & $59\pm 19$ & $12\pm 5$ & $13\pm 3$  \\
Hodge 15 & $[$H74$]$ 15 & $18$ & $4.6$ & $2.1\pm 0.8$ & $751\pm 176$ & $245\pm 89$ & $<9$ & $10\pm 3$  \\
SMCB2 South\footnotemark[1] &  & $15.4, 10.9, 15.0$ & $3.7,4.7,4.5$ & $18.8\pm 8.9$ & $1660\pm 500$ & $422\pm 211$ & $114\pm 23$ & $37\pm 10$ \\
SMCB2 North\footnotemark[2] & & $19.8, 11.4$ & $4.6,2.9$ & $21.5\pm 5.1$ & $1505\pm 280$ & $278\pm 125$ & $40\pm 10$ & $24\pm 6$ \\
SMCB2\#5 & $[$RLB93$]$ SMC-B2 3 & $15.0$ & $4.2$ & $<3.7$ & $172\pm 116$ & $57\pm 44$ & $<9$ & $4\pm 2$ \\
\end{tabular}
\end{table*}

\subsection{Origin of the 870$\mu$m emission}\label{sec:origemi}

At long wavelength like 870$\mu$m, the broad-band flux densities determined from the maps may contain non-negligible contributions by thermal free-free continuum emission and CO line emission.

\subsubsection{Free-free contribution\label{sec:ff}}

In \citet{Bot:2007yq}, estimates of the free-free emission and of the CO line at 1.2mm were found to be negligible contributions to the measured fluxes. 
Even if we expect this to be the case at 870$\mu$m, it is safe to quantify the effect here too. Furthermore, the free-free estimates obtained in \citet{Bot:2007yq} were derived from the \citet{PFR+04} radio catalog, which had integrated fluxes in regions that could be different in size with respect to the molecular clouds we are sampling, and when no source was present in the catalog it was not clear whether this was a non-detection or a non-observation. In this section, the use of the radio maps directly enables us to derive much more accurate estimates of the free-free contribution to the (sub-)millimeter emission for the regions we are interested in. 

All regions observed at 870$\mu$m and in CO(1-0) also show radio emission that is detected at 4.8 or 8.64~GHz. A few regions (e.g. LIRS49) show an inverted spectrum (which could be due to the different spatial filtering between the observations at the two frequencies). But in most cases, the radio spectral index is flat, consistent with an origin of the emission that is mostly free-free.  Indeed, for free-free emission, the flux density scales as $0.95-0.16\ln \nu$ with $\nu$ in GHz \citep{Reynolds:1992lr}, and is fitted to the radio emission detected at the highest frequency (plain black line in Fig. \ref{fig2}). Comparing this scaling with the emission at 870$\mu$m and 1.2~mm, the free-free contribution to the (sub-)millimeter emission is not significant (we estimate it to be 1 to 8\% of the 870$\mu$m emission and remove it from the observed LABOCA and SIMBA fluxes).

\subsubsection{CO(3-2) line contribution\label{sec:co32}}

The J=3-2 CO emission contribution to the 870$\mu$m broad band emission is computed using the measured CO(3-2) line, when available, from \citet{Nikolic:2007gd}, or otherwise, by applying a high CO(3-2)/CO(2-1) ratio\footnote{we take the ratio of the velocity integrated intensities in K.km.s$^{-1}$} of 2 to the CO(2-1) observed emission. Using a Jy/K conversion factor of 8.4 and a bandwidth for the LABOCA bandpass of 60GHz, we find that the CO(3-2) line contribution to the observed emission is insignificant in all clouds (it represents at most 0.3\% of the emission at 870$\mu$m). 

The 870$\mu$m and 1.2mm fluxes in each cloud are then dominated by dust continuum emission.

\subsection{Method\label{sec:meth}}

In Sec. \ref{sec:ff} and \ref{sec:co32} we showed that the emission detected by LABOCA at 870$\mu$m is dominated by dust emission. The method to determine hydrogen masses from dust emission is then similar to the one described by \citet{Bot:2007yq}. In short, the hydrogen column density in each molecular cloud can be deduced from:
\begin{equation}
N_H=\frac{I_{870}}{\epsilon^H_{870}B_{870}(T_{dust})}
\end{equation}
where $T_{dust}$ is the dust temperature, $\epsilon^H_{870}$ is the emissivity of dust per hydrogen atom at 870$\mu$m and $B_\lambda$ is the Planck law. In order to deduce dense gas masses from the dust emission detected with LABOCA, we therefore have to know the dust temperature in the clouds of the SMC and the dust emissivity at 870$\mu$m.

\subsubsection{Dust temperatures and emissivity index}

In \citet{Bot:2007yq}, it was assumed that the spectral dependance of this dust emission in the (sub-)millimeter is similar to the one in our galaxy ($S_\nu\propto \nu^\beta B_\nu(T_{dust})$ with $\beta=2.0$) with a single temperature of 15~K, typical of GMCs in the Milky Way. Even though the dust temperature dependance of the submillimeter emission is mild, the uncertainty on this parameter can create scatter in the mass determined for the sample. Both the dust temperature and the emissivity index in the molecular clouds of the SMC can be better constrained by comparing the LABOCA, SIMBA and Spitzer MIPS fluxes for each cloud.  

\begin{figure*}
\includegraphics[width=\textwidth]{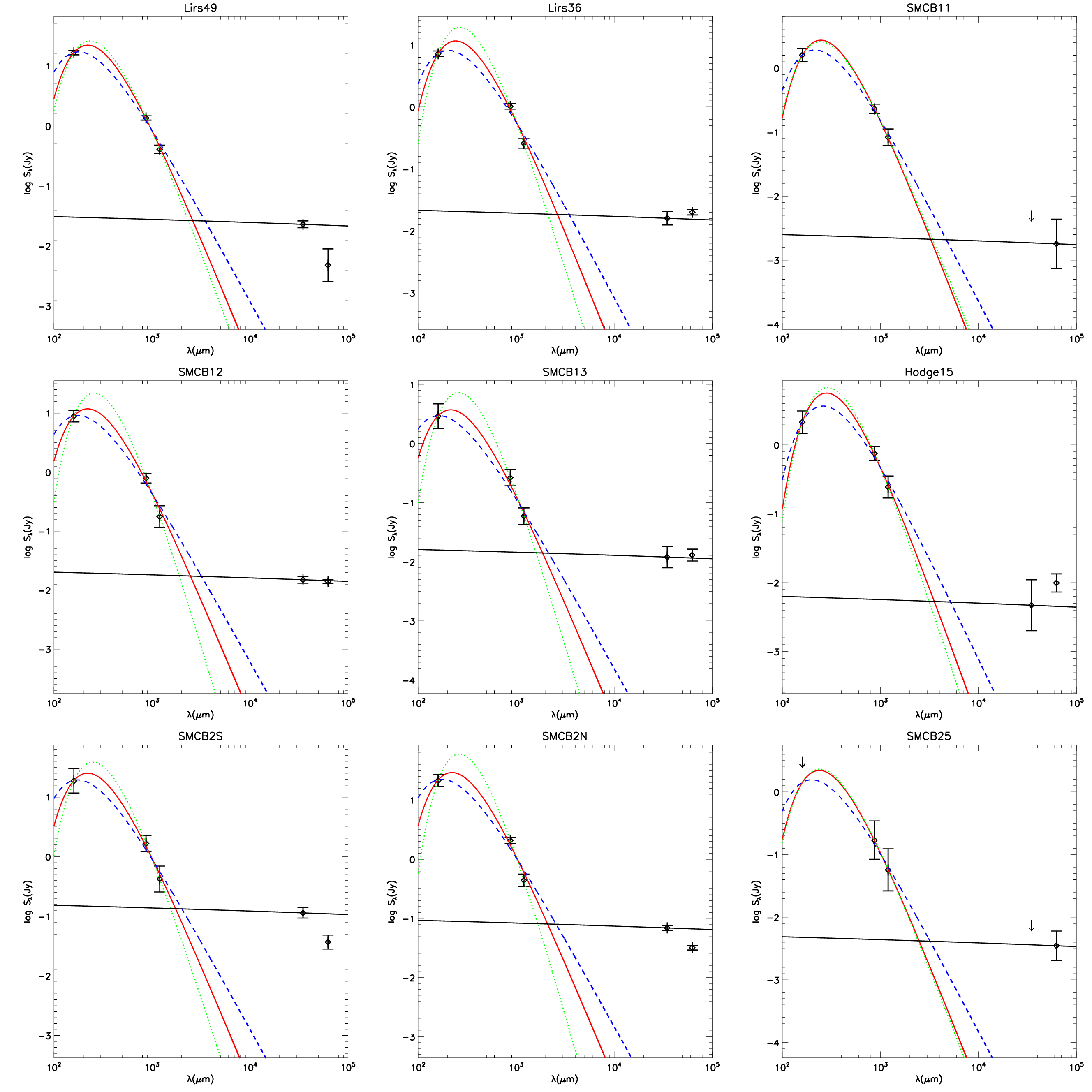}
\caption{ Spectral energy distribution from the far-infrared to the radio, observed for each cloud. The fluxes observed at 160, 870$\mu$m and 1.2mm are fitted with a single modified black body either with an emissivity index of 1 (dashed blue line) or 2 (plain red line) or with a free emissivity index (dotted green line). Radio fluxes (or upper limits) computed in the same regions are overplotted and the extrapolated free-free emission is displayed as a black line. This shows clearly that the sub-millimeter and millimeter emission is dominated by dust emission rather than free-free. \label{fig2}}
\end{figure*}

 A modified blackbody fits to the MIPS, LABOCA and SIMBA fluxes is done for each region with both the dust temperature and the emissivity index as free parameters (dashed line in Fig \ref{fig2}).
Such a fit gives values of $\beta$ between 1.9 and 4 (c.f. Tab. \ref{tab2}) and very low dust temperatures. However, high values (e.g. $\beta\sim 3$) are not believed to be physical and could be a consequence of filtering in the SIMBA data (c.f. \ref{sec:data}), even though our focus on emission peaks attempts to limit this effect. Alternatively, it could be due to a mix of temperatures within the regions we sample. 

It should be noted that since there is most probably a range of temperatures within the beam, the dust temperature determined from a single modified black body fit is an upper limit to the mean temperature. Since lower temperatures would give higher dust masses, our simple fit gives lower limits on the mass and is thus a conservative estimate to determine the dense cloud masses.

\begin{table}
\centering
\caption{Results from modified black body fits to the MIPS, LABOCA and SIMBA fluxes for each molecular cloud. The two first values correspond to a fit where both the emissivity index and the dust temeprature are free to vary, while the two last columsn corresponds to the dust temperatures determined for fixed values of the dust emissivity index ($\beta=2$ and $\beta=1$).\label{tab2}}
\begin{tabular}{l l l l l}
name & $\beta$ & $T_{dust}^{\beta}$ & $T_{dust}^{\beta=2}$ & $T_{dust}^{\beta=1}$ \\
\\
\hline
LIRS49 & $2.5\pm 0.2$ &  $11.2\pm 0.3$ &  $13.1\pm 0.2$ &$20.6\pm 0.4$ \\
LIRS36 & $3.3\pm 0.2$ & $8.5\pm 0.3$ & $12.0\pm 0.2$ & $18.0\pm 0.4$\\
SMCB1\#1 & $1.9\pm 0.3$ & $12.2\pm 0.8$ & $11.7\pm 0.3$ & $17.1\pm 0.7$\\
SMCB1\#2 &  $3.8\pm 0.4$ & $8.3\pm 0.5$ & $13.1\pm 0.4$ & $20.9\pm 1.0$\\
SMCB1\#3 & $3.8\pm 0.5$ & $8.2\pm 0.6$ & $13.5\pm 0.9$ & $22.6\pm 2.2$\\
Hodge15 &  $2.4\pm 0.5$ & $9.2\pm  0.7$ & $10.3\pm 0.4$ & $14.1\pm 0.7$\\
SMCB2 South & $3.2\pm  0.6$ & $9.3\pm 0.9$ & $13.1\pm 0.8$ & $21\pm 2$\\
SMCB2 North & $4.0\pm 0.3$ & $7.8\pm 0.3$ & $13.1\pm 0.4$ & $21.0\pm 0.9$\\
SMCB2\#5 & $2.2\pm 1.0$ & $11.7\pm 2.4$ & $12.1\pm 1.3$ & $18\pm 3$\\
\end{tabular}
\end{table}

Using the spectral energy distributions built for each cloud (Fig. \ref{fig2}) we observe that in all the molecular clouds of this study, the emission is consistent with a modified black body with a standard spectral index of $\beta=2$. Fixing the dust emissivity to this standard value, we observe low dust temperatures (12~K on average) for all the molecular clouds studied here. The temperatures range is small (from 10.2 to 13K) and consistent with a constant temperature at a 3$\sigma$ level. However, these low dust temperatures (which correspond to a radiation field intensity that would be 10 times lower than the canonical value from the solar neighbourhood \citep{MMP83}) are surprising. Indeed, some regions like LIRS36 or LIRS49 harbour active star forming regions and our regions could contain, at least partly, dust heated by the surrounding stars. However, because our study focuses on regions where CO is detected, it is possible that the dust in these regions is predominantly cold. Even though, it is surprising that dust temperature in giant molecular clouds of the SMC are lower than what is observed for molecular clouds in the Milky Way, especially since at low metallicity the dust shielding will be reduced.

The temperature and emissivity index of the dust associated with the 870$\mu$m emission is therefore still difficult to constrain and more data in the 100-300$\mu$m wavelength regime would be needed to do so.  In the following, we assume $\beta=2$ and the temperatures ($T_{dust}\sim 12K$) determined in such case from the fitting procedure, but we also compute lower limits on the mass estimates using a low value of the emissivity index: $\beta=1$. A fit to the spectral energy distributions with $\beta=1$ (dashed blue line in Fig. \ref{fig2}) leads to dust temperatures of $T_{dust}\sim19K$ (last column in Tab. \ref{tab2}). In the following, we will therefore also compute lower limits on the molecular cloud masses using $T_{dust}=19K$.

\subsubsection{Dust emissivity at 870$\mu$m \label{sec:dustemi}}

To determine gas masses from the dust (sub-)millimeter emission at 870$\mu$m it is crucial to know the dust emissivity in this wavelength regime. This quantity can be expressed as follow:
\begin{equation}
\epsilon^H_{870}=\kappa_{870}x_d\mu m_H
\end{equation}
where $\kappa_{870}$ is the absorption coefficient per unit of dust mass, $x_d$ is the dust to gas mass ratio (in the following, we adopt a dust to gas ratio 0.17 times the one in the solar neighbourhood, as determined by \citet{Bot:2007yq} from abundance studies) and $\mu m_H$ is the gas weight per hydrogen, taking into account the contribution of He. 

The long wavelength emissivity of dust in the difuse medium has been modeled and tabulated by \citet{LiD01}. At 850$\mu$m, they give $\kappa_{850\mu m}=0.4$cm$^2/$g. However, this value may not be representative of the dust properties in molecular clouds. In particular, coagulation processes can occur between grains. The large fluffy aggregates created that way have different absorption/emission properties than those in the diffuse medium. Enhancements of the dust emissivity can be attributed to coagulation processes. Such an emissivity enhancement is observed in our galaxy \citep{CBL+01,SAB+03,Bot:2007yq} and for giant molecular clouds, the dust emissivity in the (sub-)millimeter regime is enhanced by a factor 2-3 with respect to the diffuse medium. 

In order to determine the dust opacity appropriate for the GMCs probed by our observations, we chose to estimate it directly from sub-millimeter observations in a Galactic molecular environment. Without further knowledge, we make the assumption that the molecular ring in our Galaxy can be used as a reference for the giant molecular clouds of the SMC. In particular, this assumes that dust evolution (like coagulation) occur in the same way in SMC GMCs as in our galaxy. 

The method used is similar to the one by \citet{Bot:2007yq}: a map of the dust emission in our Galaxy at 870$\mu$m is created using FIRAS data. 
In the molecular ring, the 870$\mu$m dust emission correlated with H{\sc i} is substracted from the observed emission using the correlation observed at high galactic latitude. The remaining emission correlates with the CO intensity \citep{DHT01} and a linear fit on the correlation gives:
\begin{equation}
\frac{I_{870\mu m}}{W_{CO}}=0.251\pm 0.003 MJy/sr.(K.km/s)^{-1}.
\end{equation}
We checked that in individual GMCs in our solar neighbourhood, this ratio is similar ($\sim 0.26\pm 0.18$-- the uncertainty here reflects the dispersion between clouds) and is therefore characteristic of the dust emission in GMCs in the Milky Way.  Assuming a standard conversion factor between the CO intensity and the molecular gas column densities ($X_{CO}=N(H_2)/W_{CO}=1.8\cdot 10^{20}$mol.cm$^{-2}$ (K. km/s)$^{-1}$ \citep{DHT01,GCT05}), a dust temperature in galactic GMCs of 15~K as in \citet{Bot:2007yq}, we can deduce a dust emissivity at 870$\mu$m for the molecular ring. By multiplying this value with the SMC dust to gas ratio, we obtain a dust emissivity suitable for SMC GMCs:
\begin{equation}
\epsilon_{870\mu m}^H(H_2, \mathrm{SMC})=(3.94\pm 0.05)\times 10^{-27} at^{-1}.cm^{2}.
\label{eqepsi}\end{equation}
This emissivity of dust in molecular regions corresponds to an opacity $\kappa_{870}=1.26\pm0.02 cm^2/g$. 

The uncertainties quoted above for the different values correspond to formal uncertainties from the fitting procedures and do not reflect the intrinsic scatter in the 870$\mu$m-W$_{CO}$ correlation, nor the systematic uncertainty due to the assumptions which are by far the dominant source of error. For example, it is unclear whether "coagulation" of dust grains occurs in the SMC and this will be discussed further in Sec. \ref{sec:discusdust}. However, within our current knowledge, most systematic effects will increase the mass estimated from the dust emission. 

Using eq. \ref{eqepsi} and a dust temperature between 12 and 19~K, we can therefore determine gas masses directly from the observed 870$\mu$m dust emission continuum in the CO detected GMCs of the SMC.

\section{Results \label{sec:results}}

\begin{figure}
\includegraphics[width=0.5\textwidth]{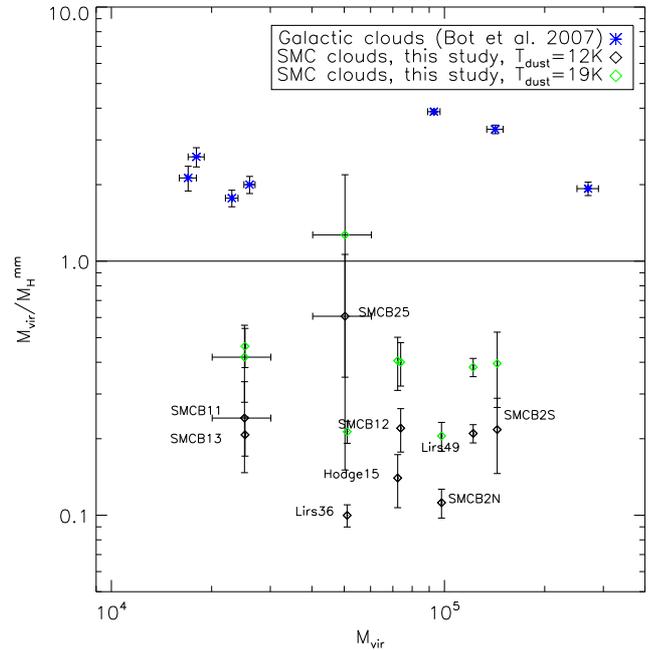}
\caption{ Comparison between the two molecular mass estimates: masses deduced from sub-millimeter dust emission and virial masses. The mass deduced from the LABOCA emission in the SMC are shown for a dust temperature as estimated  from the SED fitting with $\beta=2$ ($T_{dust}\sim 12$K, black diamonds) and for a dust temperature of 19~K as an upper limit (green rhombs). Galactic GMCs studied with the same method are shown for comparison (blue stars). In the SMC, the molecular mass deduced from the sub-millimeter are systematically larger than the one deduced from the virial theorem applied on CO data, even with an upper limit on the dust temperature, in opposition to what is seen for equivalent galactic clouds.   \label{fig3}}
\end{figure}

Gas masses for all the GMCs of the south west region of the SMC observed in CO are deduced from the LABOCA sub-millimeter emission and are summarized in Tab. \ref{tab3}. These masses are compared to the masses deduced from the application of the virial theorem on the CO line width and radii (c.f. Tab \ref{tab3} and Fig. \ref{fig3}). For comparison purposes, we took from \citet{Bot:2007yq} the masses determined from dust millimeter emission and the virial masses for a reference sample of similar molecular clouds in the solar neighbourhood.

For self-gravitating clouds, the velocity dispersion of the CO clumps balances the gravitational pressure and the virial masses should be similar to the cloud masses deduced from dust emission. However,  we observe that the masses of the GMCs as deduced from dust emission are systematically larger than the virial masses determined for the same clouds. Our study with LABOCA data at 870$\mu$m therefore confirms the results obtained by \citet{RBR+04} and \citet{Bot:2007yq} with SIMBA data. We find a median mass ratio $M_{vir}/M_H^{mm}=0.21$ (0.41 for the upper limit on the dust temperature), contrasting with the mass ratio obtained for equivalent clouds in the Milky Way, which is always above 1.

\begin{table}
\begin{minipage}[t]{\columnwidth}
\centering
\caption{ mass estimates for the individual GMCs observed in the south-west region of the SMC. The first two columns give the mass estimates computed from the LABOCA 870$\mu$m emission for dust temperatures of 12~K and 19~K respectively, while the third column lists the virial masses obtained from the CO line widths and the cloud radii.\label{tab3}}
\begin{tabular}{c c c c}
name &$M_H^{870\mu m}$(12~K) & $M_H^{870\mu m}$(19~K) &   $M_H^{vir}$\\
& $10^{4}M_\odot$ &  $10^{4}M_\odot$\\
\hline
LIRS49 & $56\pm 5$ &  $31\pm 3$& $12.2\pm 0.5$\\
LIRS36 & $50\pm 5$ & $23\pm 2$ & $5.1\pm 0.5$\\
SMCB1\#1 & $12\pm 2$ & $5\pm 1$ &$2.5\pm 0.2$\\
SMCB1\#2 & $33\pm 7$ &  $18\pm 4$ &$7.4\pm 1.4$\\
SMCB1\#3 & $11\pm 4$ & $6\pm 2$ & $2.5\pm 0.2$\\
Hodge15 & $50\pm 12$ &  $17\pm 4$ &$7.2\pm 0.4$\\
SMCB2 South & $65\pm 22$ &  $35\pm 12$&$14.4\pm 2.3$\\
SMCB2 North & $85\pm 11$ & $46\pm 6$ &$9.8\pm 1.9$\\
SMCB2\#5 & $8\pm 6$ & $4\pm 3$ & $5.0\pm 0.3$\\
\end{tabular}
\end{minipage}
\end{table}

The most straightforward interpretation of the mass discrepancy we observe in the SMC is then that the CO linewidth do not measure the full velocity distribution of the gas. There are two independent reasons for why this could be true.

The CO emitting regions may be present only in the interior of the molecular clouds, in regions shielded from the UV radiation field \citep{LLD+94,Glover:2010uq}. If the (unresolved) CO emitting clumps are not evenly distributed in the region detected by the CO observations, the observed line width might not reflect the full velocity dispersion in the cloud.

The CO line could come from the densest regions created by shocks in a turbulent medium \citep{Klessen:2009jt}. Simulations in a turbulent interstellar medium show that these dense regions with CO might not always be the most shielded ones \citep{Glover:2009hb}. In this case, the velocity dispersion in the CO emitting regions would not be representative of the velocity dispersion of the molecular gas located also in lower density regions. The CO line width would then underestimate the velocity dispersion of the molecular gas due to its preferential location in the densest and less turbulent regions. Such a difference has been observed between the CO and HI line width in high latitude clouds in the Solar Neighbourhood on a few parsec scale \citep{Barriault:2010fk}, but it is unclear if such a difference would exist on larger scales where many shocks at different locations of the cloud must be occuring. 

Due to the low metallicity of the SMC, a Galactic analog to understand the CO emitting clouds in the SMC may be denser molecular environments. Indeed, if the H$_2$ abundance varies as the product of the mean gas number density and the metallicity \citep{Glover:2010uq}, the motions observed in CO and the mass observed in the dust continuum can be compared to the velocity of molecular cores and the relation to their enveloppes in our Galaxy. 
In "extinction super cores",  \citet{kirk:2007kx} observed that the inner N$_2$H$^+$ or C$^{18}$O starless cores show a velocity dispersion that is systematically lower than the one expected for a gravitational support of the region in extinction. The motions between the densest molecular entities are hence not sufficient to support the enveloppes against gravity, similarly to what we observe in the giant molecular clouds in the south-west of the SMC.

\section{Discussion}

Our analysis of the two mass estimates was done in a way to avoid as much as possible the known caveats (CO-devoid envelope, enhanced dust emissivity): fluxes are extracted from regions restricted to where CO is observed, we attempt to remove the extended emission surrounding the CO peaks and the dust emissivity is taken for molecular environments. The impact of such effects on the masses deduced is checked in Appendix \ref{append}. Here we discuss the possible remaining biasses that could hamper the analysis and result in the mass discrepancy. We also discuss alternative interpretations for the mass discrepancy we observe.

\subsection{Unresolved CO clouds and geometrical effects\label{sec:geom}}

In the scenario where CO is confined inside a large H$_2$ envelope, then the dust emission will trace the total cloud (CO-emitting cloud+envelope). In order to minimize the effect of the CO-devoid envelope on the dust emission, we performed a local estimate of the extended emission around the CO-detected region and removed it from the measured (sub-)millimeter fluxes. However, the effectiveness of this removal depends on the size of the envelope and on the density profile of the cloud.

For the same configuration, the virial mass is obtained from the CO velocities and the radius of the CO cloud. This radius may be overestimated from the observations if the CO clumps are not resolved and the virial masses will be overestimated. 
The dimension differences between the dust emitting region, the CO emitting region and the area deduced from the observations (i.e. the aperture) may then still bias the ratio M$_{vir}$/M$_{mm}$ in either direction (enhancing or decreasing the mass ratio). 

The exact magnitude of such biases is a complicated function of the real structure of the cloud (e.g. density), the definition
of the CO cloud and aperture, and the extent of any CO-free dusty envelope surrounding the cloud. We tested the magnitude of a geometric bias for a range of density profiles with a simple model. In this model, the CO cloud and the envelope are spherical. A circular aperture defines the radius that enters the virial mass calculation, while also defining the impact of the background subtraction to the mass deduced from the dust emission. We find that for a simple model in which dust emerges from a cloud twice the size
of the CO cloud and the aperture covers the portion of the cloud not showing CO emission, the bias almost never exceeds $\sim$30\%.
However, if the CO cloud is actually substantially smaller than the values quoted in Table \ref{tab1} (e.g. CO clumps are not resolved at the SEST resolution), if the dust envelope is very large compared to the CO cloud, or if the aperture extends too far from the surface of the cloud, these effects may be larger. Conversely, for shallow density profiles and aperture definitions too close to the surface of the cloud, it is possible to over-correct for a dusty envelope, meaning that our measurement of $M_{mm}$ would in fact be too low and exacerbating the difference between $M_{vir}$ and $M_{mm}$. This effect has been explored in more details for the N83 region in the Wing of the SMC  \citep{Leroy:2009rt} and could play a role in the mass discrepancy despite our attempt to limit the mass comparison to the CO peaks of the molecular clouds.

To further assess these biases, we performed the mass calculations without the extended emission subtraction and with deconvolved CO radii. The results are presented in Appendix \ref{append}. It shows that, as far as can be tested it with the current observations, both effects seem to enhance the mass discrepancy.

Observations with better resolution and comparisons to models and simulations of cloud structure should offer the chance to better constrain whether geometric effects can drive the observed difference. Here we note the possible concern and that we have done our best to correct for it given the data available (via the extended emission fitting procedure). 

\subsection{The dust to gas ratio and the dust emissivity \label{sec:discusdust}}

In order to compute molecular cloud masses from the dust emission, we assumed that the dust properties in the Small Magellanic Cloud are similar to the one we observe in our Galaxy, except for the lower dust content with respect to the gas due to the metallicity. Hence, one possibility to understand the mass discrepancy we observe is to question our understanding of dust emission in the SMC. 

Unfortunately, our current knowledge of dust grains in the SMC is limited and the dust emission in the SMC could be different than in our Galaxy. Indeed, \citet{Israel:2010fk} observed a millimeter emission excess in the Small Magellanic Cloud, on the scale of the galaxy, that is not explained by current standard models of dust emission \citep{Bot:2010uq}. The nature of this excess remains unclear but might be associated to spinning dust emission. Given the fact that dust emission in the SMC is not well understood on a global scale, the interpretation of a similar excess emission on the scale of individual molecular clouds is therefore precarious. We acknowledge this uncertainty but do our best within the current knowledge on dust properties in the SMC, i.e. we assume that large dust grains at thermal equilibrium do most of the sub-millimeter dust emission.

Biases due to our uncertainties on the dust temperature and emissivity index (note that these quantities are not independent) were assessed through the use of an extreme value $\beta=1$ giving $T_{dust}\sim 19$~K for a fit of the observed far-infrared to millimeter SEDs. In Fig. \ref{fig3}, we observe that the lower limit on the molecular cloud masses obtained with these parameters are still larger than the virial masses.

In Sec. \ref{sec:dustemi}, we assumed that the dust emissivity is equivalent in the SMC to what is observed in GMCs of the Milky Way, and scaled it accordingly to the dust-to-gas ratio difference between the two galaxies.  However, there is no ground based evidence that this is the case. First, the knowledge of the dust-to-gas ratio in the SMC is quite uncertain. Second, the grain opacity could be different. In the following, we discuss how these two quantities could vary.

The dust-to-gas ratio estimate we use is based on abundance measures toward the star AV 304 \citep{RVT+03}, using galactic depletions computed from \citet{SM01a,SM01b}. These abundances are consistent with all  HII region abundances published for the SMC \citep{VV02,Perez-Montero:2005jw,Lebouteiller:2008ao} expect for one (N13, for which a close to solar O/H abundance has been deduced) so that we believe our metallicity estimate is accurate for the SMC. The depletion factors could be higher than in the Milky Way, enhancing the dust-to-gas ratio and decreasing the mass discrepancy. However, in order to compensate the mass discrepancy we observe with a change of the dust-to-gas ratio, all metals would have to be depleted onto dust grains. It therefore seems unlikely that a change of the dust-to-gas ratio could explain the mass discrepancy we observe.

Grain coagulation can make the sub-millimeter dust emissivity larger than the \citet{LiD01} value referenced for the diffuse medium. An enhancement (a factor of 3) is observed in Galactic GMCs. We assumed that the same enhancement factor applies to the SMC clouds in our study. Could this enhancement be larger in the SMC and explain the mass discrepancy? Taking an upper limit on the dust temperature, we observed in the last section that the virial mass for the SMC GMCs is at least 2 to 5 times smaller than the mass deduced from the dust millimeter emission. In order to get $M_{vir}=M_H^{mm}$, the dust emissivity in the SMC would then need to be at least 2 to 5 times larger than the value we use.
 We note that such an enhancement would lower dust temperatures by a factor $N^{\frac{1}{4+\beta}}$ (where $N$ is the factor of emissivity enhancement), since the dust grains would radiate more efficiently. Coagulation could therefore explain both the mass discrepancy and the low dust temperatures we observed when fitting the spectral energy distributions with an emissivity index $\beta=2$.

\citet{Voshchinnikov:2006ek} analyzed the influence of grain porosity on the dust properties. They show that as the porosity increases, the millimeter mass absorption increases and the dust temperature decreases. Using their results, we see that the mass absorption coefficient for a porous composite grain (amorphous carbon+astronomical silicate) can be up to 22 times the \citet{LiD01} value at 1~mm. Therefore, in theory the mass discrepancy between the virial theorem and dust estimates could be due to a larger fraction of porous grains or a larger porosity of the grains in the SMC GMCs than in the Milky Way.

However, since grains coagulation is proportional to the density of dust particles, one would expect less coagulation of dust grains in the SMC than in our galaxy, due to the lower dust content. Indeed, recent models of the regulation of the grain size distribution by shattering and coagulation \citep{Hirashita:2008zr} suggest that, for a MRN grain size distribution \citep{MRN77}, the importance of both these processes is significantly reduced at low metallicities like the one in the SMC. But the grain size distribution in the SMC seems significantly different than the one in our galaxy: the differences between the SMC and Milky Way diffuse ISM extinction curve and infrared spectral energy distribution indicate a larger fraction of the dust mass in small dust grains \citep{WD01, BBL+04, Bernard:2008zr}. This enhancement in small dust particles is therefore likely to also impact the grain coagulation process. Using the dust grain size distribution for the SMC and the Milky Way (for R$_V=3.1$) given by \citet{WD01} and a simple coagulation model (using the collision rates due to turbulence from \citet{Draine:1985ul} and assuming that grain stick when they collide), we observe that for a given total dust mass, small grains ($10\AA< a<0.1\mu$ m) could coagulate twice more quickly on big grains than in our galaxy. 
The enhancement of small grains in the grain size distribution of the SMC therefore seems to favor coagulation processes while the low metallicity of the SMC would penalize it. More detailed models would be needed to quantify the probability of large porous aggregates in the SMC. Such models are beyond the scope of the current paper. It seems unlikely at present that large porous aggregates would be much easier to form in the SMC than in our galaxy. But the type of aggregates formed in the SMC (and hence their porosity) could be different than the one in the Milky Way due to the different size distribution and this could also potentially affect the dust emissivity.

We can not completely exclude at present that the mass difference we observe between the two methods (virial theorem and dust millimeter emission) is due to a higher sub-mm emissivity than that measured in  Galactic GMCs. The presence of large porous grains in the SMC through coagulation would indeed explain both the mass difference and the low temperatures observed. But our current understanding of dust in the SMC does not favor more coagulation in this environment than in our galaxy. 

\section{Conclusion\label{sec:conclusion}}

New LABOCA observations at 870$\mu$m were obtained in the south-west region of the Small Magellanic Cloud. All the GMCs observed previously in CO are detected. We build spectral energy distributions by comparing the fluxes measured at 160$\mu$m, 870$\mu$m, 1.2~mm, 3 and 6~cm for each cloud. We observed that the sub-millimeter and millimeter emission, in all clouds of this study, is always dominated by dust emission. Fitting the spectral energy distribution with a modified blackbody with a standard $\beta=2$ gives surprisingly low dust temperatures $T_{dust}\sim 12$K. Given the uncertainties, we also took $\beta=1$  (giving fitted temperatures around $T_{dust}\sim 19$K) to deduce lower limits on the molecular cloud masses obtained from the dust emission. Using a dust emissivity at 870$\mu$m that we determine in molecular regions of our galaxy and scaled to the dust-to-gas ratio of the SMC, mass estimates for the molecular clouds observed (as well as lower limits on these masses) are derived from the 870$\mu$m dust emission and compared to virial masses deduced from CO observations.

We confirm the discrepancy between the virial masses and the one deduced from dust emission that was observed previously in the molecular clouds of the SMC. This discrepancy has now been observed at two different wavelength with two different instruments and as such can no longer be attributed to instrumental effects. Furthermore, uncertainties on the origin of the sub-millimeter emission have been assessed (free-free contribution, CO line contribution). The mass ratio $M_{vir}/M_H^{mm}$ observed in the SMC is clearly below one, even for conservative assumptions on the dust emissivity, dust temperature, etc. In contrast, for similar clouds in the solar neighbourhoods the observed values are systematically above 1. The direct interpretation of this result is that the CO line widths do not trace the full velocity dispersion of the gas in giant molecular clouds in the SMC. This could be due to the turbulent nature of the intersellar gas and the formation of CO in the densest structures.

This low mass ratio in the SMC is unexpected for large molecular clouds and will have to be studied further in order to assess its origin. The dust properties in the SMC could be different than in our galaxy and we can not completely exclude that an enhanced grain coagulation in the SMC could increase the dust emissivity and lead to an overestimate of the true mass by the dust millimeter emission,  Herschel observations will give more informations in that respect by complementing the wavelength coverage. The cloud+envelope geometry could also play a role in enhancing or decreasing the mass discrepancy and higher resolution observations will be needed to assess this uncertainty. 

Further investigations are necessary in order to better understand the origin of this discrepancy, whether it is a general property of clouds in low metallicity galaxies and what it tells us about the molecular clouds properties. This will be better done with Herschel, which will be sensitive to dust emission on all angular scales and will provide a better spectral sampling of the far-infrared and sub-mm emission. Also, spectroscopic observations of the C$^+$ line with Herschel will probe the extended enveloppes of the molecular clouds and enable us to test whether the C$^+$ linewidth are comparable to the one observed in CO (i.e. test whether CO velocities trace the motions inside the molecular clouds).

\begin{acknowledgements}
M.R wishes to acknowledge support from FONDECYT grant No 1080335. M.R. is supported by the Chilean {\sl Center for Astrophysics} FONDAP No. 15010003.M.R ,F.B, and C.B have been supported by  the ECOS-CONICYT C08U03 project.
\end{acknowledgements}

\bibliographystyle{aa}
\bibliography{../aamnem99,../../../biblio}

\begin{appendix} 
\section{Systematics \label{append}}
\subsection{Effect of the local background subtraction on the masses derived from the sub-mm emission\label{append1a}}

For this study, we tried to minimize the effect of CO-devoid enveloppe on the dust emission by performing a local background subtraction. This option was chosen to be conservative in order to compute masses from dust sub-millimeter emission that are comparable to virial masses obtained from CO observations. However, it can lead to complicated uncertainties as discussed in Section \ref{sec:geom}. 

In this appendix, we test the impact of the background subtraction on our results by redoing the analysis on the molecular clouds in the south-west region of the SMC, but without local background subtraction. Instead, we took the same background region for all clouds in all maps at the location 00:46:02.85 -73:19:26.5 (J2000). The fluxes obtained from 160$\mu$m to 6~cm are given in Table \ref{taba1}. On average, they are 1.5 times higher at 870$\mu$m. All further steps of the analysis were then reproduced in the same manner as described in Sect. \ref{sec:meth} and the masses obtained are given in Tab. \ref{taba3}. Doing so, we find a median mass ratio $M_{vir}/M_H^{mm} =0.14$ (0.23 for the upper limit on the dust temperature), i.e. 1.5 times lower than with the local background subtraction. This confirms that our attempt to minimize the effect of an extended enveloppe by subtracting a local background to the emission observed in each cloud, decreases the mass discrepancy observed the submillimeter emission and the virial masses. 

\begin{table*}
\centering
\caption{Same as Tab. \ref{tab1} but the radii correspond to deconvolved CO radii and the fluxes have been extracted without the local background subtraction. When the observed radius of the cloud was lower than the beam, the deconvolved radius is then set to NaN and we use the observed CO radius to compute masses instead. \label{taba1}. }
\begin{tabular}{l l l  l l l l l l l}
name &  Radius & $S_{160\mu m}$ &$S_{870 \mu m}$ &  $S_{1.2mm}$ & $S_{3cm}$ & $S_{6cm}$ \\
 & pc & Jy &mJy & mJy  &mJy & mJy  \\
\hline
LIRS49 & 10.6 &$39.5\pm 3.4$ & $2241\pm 266$ & $457\pm 91$ & $36\pm 5$ & $36\pm 4$  \\
LIRS36 & 13.4 & $24.5\pm 4.6$ & $1551\pm 140$ & $462\pm 90$ & $80\pm 30$ & $74\pm 21$  \\
SMCB1\#1  & 4.8 &$3.9\pm 0.9$ & $507\pm 120$ & $215\pm 63$ & $<10$ &$<5$  \\
SMCB1\#2 & 9.5 & $10.1\pm 2.1$ & $1252\pm 185$ & $265\pm 71$ & $21\pm 6$ & $24\pm 3$  \\
SMCB1\#3  & NaN &$5.4\pm 1.4$ & $589\pm 99$ & $69\pm 19$ & $12\pm 4$ & $13\pm 3$  \\
Hodge 15 & 12.5 & $0.68\pm 0.35$ & $467\pm 83$ & $176\pm 43$ & $2.3\pm 1.2$ & $5\pm 2$  \\
SMCB2 South\footnotemark[1] & 8.4, NaN, 7.2& $45\pm 9$ & $2603\pm 557$ & $917\pm 187$ & $114\pm 20$ & $105\pm 25$ \\
SMCB2 North\footnotemark[2] & 14.8, NaN &  $43\pm 6$ & $2137\pm 330$ & $577\pm 165$ & $79\pm 15$ & $87\pm 10$ \\
SMCB2\#5 & 7.6 &$6.3\pm 1.5$ & $170\pm 116$ & $71\pm 44$ & $27\pm 17$ & $27\pm 17$ \\
\end{tabular}
\end{table*}

\subsection{deconvolved CO cloud radii}

We have used the  sizes of the CO clouds as reported by Rubio et al. 1993b
and Lequeux et al. 1994. Their sizes are the observed sizes which
were not de-convolved by the beam size as the CO clouds most often showed larger sizes
than the SEST beam. If one de-convolves their observed sizes assuming the standard
aproximation of a source size smaller than the beam size, then the cloud sizes become smaller
than the observed sizes published and used in the main text of this paper. The raddi obtained for the deconvolved CO clouds are given in Table \ref{taba1}. For a 43"arcsecond beam the de-convolved
sizes are smaller than the observed sizes by a fraction of 0.3 to 0.8.  In some cases, when the CO map had been fully sampled, the observed CO radius can be already smaller than the beam size. In this case, the deconvolved radius is marked in Table \ref{taba1} as 'NaN' and we use the observed radius to compute virial masses. Since deconvolved radii are smaller than the observed ones, this results in smaller virial masses for the clouds by a similar fraction (see Table \ref{taba3}). Thus the difference between the gas mass derived from the dust emission becomes even larger.

Combining the effect of deconvolved radii with the lack of the enveloppe subtraction (from Appendix \ref{append1a}), we obtain a median mass ratio $M_{vir}/M_H^{mm}=0.10$ (0.16 for the upper limit on the dust temperature), i.e. twice lower than the one presented in the main text of this paper. This confirms that our different attempts  to minimize known caveats like the presence of a  CO-devoid enveloppe and the resolution of the CO observations, decrease the mass discrepancy. The mass discrepancy observed in the south-west region molecular clouds could then be at least twice larger.

\begin{table}
\begin{minipage}[t]{\columnwidth}
\centering
\caption{ Gas masses deduced from the sub-millimeter dust emission (as Tab. \ref{tab3} but without local background subtraction) and virial masses deduced with the deconvolved CO radii.\label{taba3}}
\begin{tabular}{c c c c}
name &$M_H^{870\mu m}$(12~K) & $M_H^{870\mu m}$(19~K) & $M_H^{vir}$ \\
& $10^{4}M_\odot$ &  $10^{4}M_\odot$ & $10^4$M$\odot$\\
\hline
LIRS49 & $75\pm 9$ &  $50\pm 6$ & $7.7\pm 0.5$\\
LIRS36 & $57\pm 5$ & $34\pm 3$ & $3.7\pm 0.5$\\
SMCB1\#1 & $26\pm 6$ & $12\pm 3$ &$0.9\pm 0.2$ \\
SMCB1\#2 & $56\pm 9$ &  $28\pm 4$ & $4.3\pm 1.4$\\
SMCB1\#3 & $21\pm 4$ & $13\pm 2$ & $2.5\pm 0.2$\\
Hodge 15 & $39\pm 7$ & $11\pm 2$ & $5.0\pm 0.4$ \\
SMCB2 South & $95\pm 21$ &  $57\pm 13$ & $9.5\pm 2.3$\\
SMCB2 North & $72\pm 12$ & $47\pm 8$ & $7.8\pm 1.9$\\
SMCB2\#5 & $4\pm 3$ & $3\pm 2$ & $2.6\pm 0.3$ \\
\end{tabular}
\end{minipage}
\end{table}

\section{Sub-mm dust emission outside the CO clouds\label{appendb}}

\begin{figure}
\includegraphics[width=0.5\textwidth]{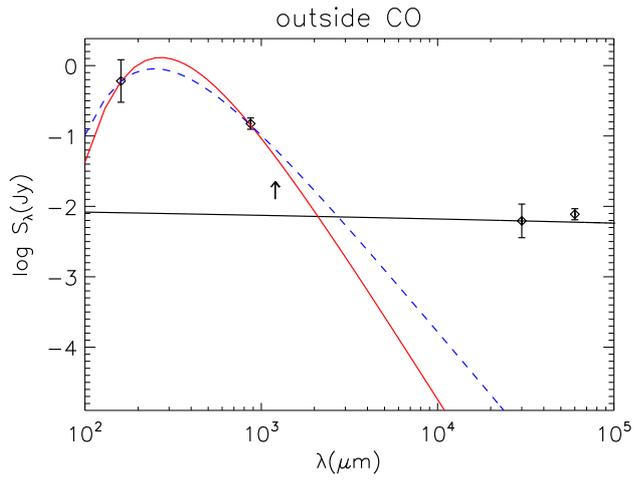}
\caption{ Spectral energy distribution from the far-infrared to the radio, observed for the region outside the molecular clouds observed in CO. The fluxes observed at 160, 870$\mu$m are fitted with a single modified black body either with an emissivity index of 1 (dashed blue line) or 2 (plain red line). Radio fluxes computed in the same regions are overplotted and the extrapolated free-free emission is displayed as a black line.  \label{fig1_ab}}
\end{figure}

The LABOCA 870$\mu$m map of the south-west region presented in this study shows extended sub-millimeter emission well outside the molecular clouds observed in CO. In this appendix, we present a short analysis of this extended dust emission outside the CO peaks. We build the spectral energy distribution from 160$\mu$m to 6cm by extracting fluxes at each wavelength for a region defined as being observed at all wavelengths and being outside the CO peaks studied. This region hence defined corresponds to a solid angle of $1.00787\times 10^{-7}$ sr. For the extended emission observed in this region, the spatial filtering in the SIMBA 1.2mm observations is no longer negligible and the flux obtained at 1.2mm is considered as a lower limit. The spectral energy distribution obtained is shown in Fig. \ref{fig1_ab}.

A free-free emission model is fitted to the 3cm radio flux and extrapolated to 160 and 870$\mu$m. The free-free contribution is then removed from the observed flux at 160 and 870$\mu$m before fitting a modified black body spectrum with a spectral index of 1 (blue dashed line) or 2 (plain red line). The dust temperatures deduced are $T_{dust}^{\beta=1}=14.8$K and  $T_{dust}^{\beta=2}=10.8$K, respectively. The dust temperature observed in the region surrounding the CO peaks is therefore similar to those observed in the CO-detected regions in our observations.

The hydrogen mass of this region deduced from the dust emission fit is $9.3\times 10^4 M\odot$ for a standard spectral index $\beta=2$ ($5.2\times 10^4 M\odot$ for $\beta=1$). This is comparable to the mass, deduced from the sub-mm dust emission, of the smallest CO-detected molecular clouds of this study.

Using the H{\sc i} observations of the SMC from \citet{SSD+99}, we estimated the neutral hydrogen mass in this region to be $2.5\times 10^4 M\odot$, i.e. 2 to 4 times lower than the mass of the gas deduced from the dust emission. Most of the gas observed through the extended sub-mm emission outside CO-detected molecular clouds is then probably cold molecular hydrogen. This is consistent with the fact that we observe no correlation between the 870$\mu$m emission and the H{\sc i} 21 cm emission.

\end{appendix}

\end{document}